\documentclass{article}
\usepackage{arxiv}
\usepackage[numbers,sort&compress]{natbib}
\usepackage{textcomp}

\usepackage{amsmath,amsfonts}

\usepackage[hidelinks]{hyperref}
\usepackage{url}
\usepackage{xurl}

\usepackage{graphicx}
\usepackage{booktabs}
\usepackage{wrapfig}

\title{Conditional Dynamical Systems for Image Generation}

\author{%
  Lianlong Sun\thanks{Equal contribution.}\thanks{Corresponding author.} \\
  Department of Electrical and Computer Engineering \\
  University of Rochester \\
  Rochester, NY, USA \\
  \texttt{lianlongsun@rochester.edu} \\
  \And
  Chuan Liu\footnotemark[1] \\
  Department of Electrical and Computer Engineering \\
  Rice University \\
  Houston, TX, USA \\
  \texttt{cl320@rice.edu} \\
  \AND
  Yongchao Liu \\
  Department of Electrical and Computer Engineering \\
  University of Rochester \\
  Rochester, NY, USA \\
  \texttt{yongchao.liu@rochester.edu} \\
  \And
  Hui Wu \\
  Department of Electrical and Computer Engineering \\
  University of Rochester \\
  Rochester, NY, USA \\
  \texttt{hui.wu@rochester.edu} \\
  \AND
  Tony (Tong) Geng \\
  Department of Electrical and Computer Engineering \\
  Rice University \\
  Houston, TX, USA \\
  \texttt{tg62@rice.edu} \\
  \And
  Michael Huang \\
  Department of Electrical and Computer Engineering \\
  University of Rochester \\
  Rochester, NY, USA \\
  \texttt{michael.huang@rochester.edu} \\
}

\date{}
\renewcommand{\shorttitle}{Conditional Dynamical Systems for Image Generation}
\hypersetup{
  pdftitle={Conditional Dynamical Systems for Image Generation},
  pdfauthor={Lianlong Sun, Chuan Liu, Tony (Tong) Geng, Michael Huang}
}
\begin{document}

\maketitle

\begin{abstract}
Image generation is dominated by deep generative models on GPUs, whose computational and energy costs raise sustainability concerns. Emerging non-von Neumann substrates, including quantum, compute-in-memory, photonic, and thermodynamic platforms, promise greater efficiency, yet most work ports conventional neural network architectures onto them and mainly accelerates operations such as matrix multiplication. This does not fully exploit a native capability of many hardware substrates: relaxation toward low-energy states can itself compute at little cost. Building on this primitive, we develop dynamical systems for image generation based on Ising spins and Kuramoto oscillators. The internal states of both models evolve under dynamics with an explicit Lyapunov energy, and a compact decoder renders the final states as images. For conditional generation, we introduce local conditioning: programmed interactions remain fixed, and one class-dependent bias per node reshapes the landscape, avoiding additional reprogramming and calibration costs. On CIFAR-10, the Ising and Kuramoto models reach clean-FID scores of $6.40$ and $5.37$, respectively, with $4096$ nodes. With a CMOS-compatible hardware realization, the dynamical-system core is projected to cost ${\sim}110.6$~nJ per image, and the full computation, even with the decoder on a GPU, outperforms all GPU baselines in latency and energy. These results suggest that energy-descending dynamics on non-von Neumann substrates can serve as generative computation, offering an efficient path beyond GPUs.
\end{abstract}

\section{Introduction}

Image generation is a central task in machine learning, with applications in content creation, data augmentation, and design. Generative adversarial networks, diffusion models, and flow-matching models have achieved high generation quality~\citep{goodfellow2014gan,ho2020ddpm,lipman2023flow}. Their success, however, relies on increasingly capable GPU infrastructure. Generation requires substantial computation and energy, motivating interest in substrates whose native physical behavior differs from that of conventional processors~\citep{strubell2019energy,patterson2021carbon}.

Several classes of non-von Neumann substrates have been explored. Quantum annealers and Ising machines encode an objective in physical interactions and evolve toward low-energy configurations~\citep{johnson2011quantum,moy2022ring,zhang2024reviewising}. Analog compute-in-memory systems based on memristive crossbars or phase-change memory perform matrix operations where parameters are stored, reducing data-movement cost~\citep{ielmini2018inmemory,sebastian2020memory,mannocci2026achieving}. Photonic processors exploit optical propagation and interference for ultrafast computation~\citep{shen2017deep,kalinin2025analog}, while thermodynamic, spintronic, and stochastic hardware uses physical fluctuations as a native sampling resource for computation~\citep{borders2019integer,melanson2025thermodynamic}.

For machine learning, these substrates are mostly used as accelerators: the neural network keeps its GPU-oriented architecture, and selected operations are offloaded to specialized devices~\citep{shen2017deep, ambrogio2023analog,oguz2024optical}. This approach can be effective, but it does not exploit the native physical primitive, which often runs at negligible energy and latency cost. Offloading may also incur overhead and lose accuracy to programming error, quantization, parasitics, and noise, reducing part of the efficiency gain~\citep{xiao2022accuracy}.

We take a different direction: designing the model around the hardware substrate's computational primitives~\citep{momeni2025training,bosch2025local,whitelam2026generative, jelincic2026efficient}. An initial demonstration shows that a coupled-oscillator system can produce nontrivial image distributions~\citep{unconventional_ai_un0_2026}. Beyond this first step, designing and realizing generative models for such substrates raises two challenges. First, what computation should the model perform so that existing hardware can run it natively? Second, how should the image class be introduced when the interaction array is written to physical devices and is expensive to change?

\begin{wrapfigure}{r}{0.472\textwidth}
\vspace{-12pt}
    \centering
    \includegraphics[width=0.472\textwidth]{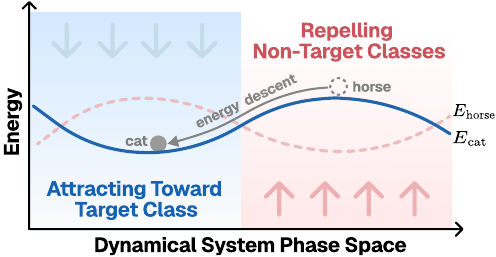}
    \caption{Generation by energy descent. Local conditioning reshapes the landscape, attracting the evolving state toward the target class.}
    \vspace{-5pt}
    \label{fig:energy-landscape}
\end{wrapfigure}

We address both challenges with one design principle: the substrate dynamics descend an explicit energy, and conditioning reshapes this energy landscape instead of reprogramming the interactions. Figure~\ref{fig:energy-landscape} illustrates the idea: when the class \emph{cat} is selected, the energy landscape attracts the state toward cats and repels it from horses. Generation becomes energy relaxation, the primitive that the hardware performs natively at little cost. For conditional generation, we introduce \emph{local conditioning}: all pairwise couplings are fixed, and class information enters only through one local bias per node, which tilts the shared landscape. Switching the class updates only these values and leaves the programmed interaction array untouched. This avoids reprogramming and calibration costs, which could otherwise dominate the cost of generation.

The design principle is not limited to one substrate. We study both continuous Ising dynamics and coupled Kuramoto oscillators and map them to Ising machines. Both descend a learned energy landscape with the local conditioning mechanism, and a compact convolutional decoder turns the final state into an image. The energy is defined over the internal state rather than pixels: energy descent produces the representation the decoder reads.

Many existing Ising machines, quantum and digital annealers, and thermodynamic, memristive, and photonic substrates already provide most of the primitives our design requires. Our results suggest that such systems could gain generative capability with modest architectural changes. Several of these substrates are CMOS-compatible, so the same dynamics could also be built as a dedicated generative core inside a GPU or as a standalone chip. For now, our goal is not to replace mature GPU-native generators, whose parameter budgets and primitives differ. Instead, we ask whether hardware built for energy minimization can generate images with its native relaxation dynamics, and at what cost. Our contributions are:

\begin{itemize}
    \item We formulate conditional image generation as dynamics that descend an explicit Lyapunov energy, with two models: continuous Ising spins and coupled Kuramoto oscillators.
    \item We introduce local conditioning, which selects the class through one constant bias per node while the pairwise interactions stay fixed, avoiding reprogramming and calibration costs.
    \item With $n=4096$ nodes, the Ising and Kuramoto models reach clean-FID scores of $6.40$ and $5.37$ on CIFAR-10, respectively, outperforming existing substrate-native baselines.
    \item We present a case study of a CMOS-compatible hardware design. The dynamical-system core is projected to cost ${\sim}110.6$~nJ per image. With a compute-in-memory decoder accelerator, the projected energy and latency of the full computation are 5.9~\textmu J and 57.4~\textmu s per image, far below those of GPU-based generators.
\end{itemize}
\newpage

\section{Method}
\label{sec:method}

We first formulate our model (Section~\ref{sec:overview}) and its two dynamics: continuous Ising spins (Section~\ref{sec:state}) and Kuramoto oscillators (Section~\ref{sec:kuramoto}). Next, we present our local conditioning mechanism, which enables conditional generation while keeping the pairwise couplings fixed across classes (Section~\ref{sec:tilt}). Finally, we describe the training procedure (Section~\ref{sec:loss}) and show how the resulting model maps onto physical hardware (Section~\ref{sec:hw_mapping}).

\subsection{Model Overview}
\label{sec:overview}

The generator maps one of $C$ class labels, $y\in\{1,\dots,C\}$, and a random initial state $\xi$ to an image,
\begin{equation}
  G(y,\xi) = \mathrm{Dec}_\psi\!\left(\Phi_\phi(y,\xi)\right),
  \label{eq:generator}
\end{equation}
where $\Phi_\phi$ returns the terminal state of a physical evolution and $\mathrm{Dec}_\psi$ is a convolutional decoder. The random initial state $\xi$ is the source of sample diversity. We consider two choices for $\Phi_\phi$: continuous Ising spins and coupled phase oscillators. In both cases, a pairwise interaction is shared by all classes, while one bias per node selects $y$.

\subsection{Continuous Ising Dynamics}
\label{sec:state}

The continuous-spin model begins from the binary Ising energy~\citep{ising1925}
\begin{equation}
  E_{\mathrm{Ising}}(s)=-\frac{1}{2}\sum_{i\neq j}J_{ij}s_i s_j-\sum_i b_i s_i,
  \qquad J_{ij}=J_{ji},\quad J_{ii}=0,
  \label{eq:ising}
\end{equation}
where the spins are discrete, $s\in\{-1,+1\}^n$, $J$ contains pairwise couplings, and $b$ contains the coefficients of the linear term. Quadratic unconstrained binary optimization problems map directly to this energy~\citep{lucas}. In Max-Cut, for example, negated edge weights become couplings, and the lowest-energy state is the best cut. An Ising machine solves such problems physically: its state settles toward low energy, and the final configuration is a good solution.

We let each spin take a continuous value, $x\in\mathcal X=[-1,1]^n$, as in most analog Ising machines~\citep{yamamoto2017coherent,afoakwa2021brim}. We define the system energy as
\begin{equation}
  E_y^{\mathrm I}(x)=-\frac{1}{2}x^\top A x-\gamma b_y^\top x,
  \qquad A=A^\top\in\mathbb{R}^{n\times n},\quad b_y\in\mathbb{R}^n.
  \label{eq:energy}
\end{equation}
The vertices of $\mathcal X$ are binary Ising configurations. The interaction $A$, which extends $J$ with a diagonal, is shared across classes. Here $b_y$ is the local conditioning vector of class $y$, and the fixed scalar $\gamma$ sets its strength.

Starting from $x(0)=\xi\sim\mathcal U(\mathcal X)$, the relaxation is
\begin{equation}
  \dot{x}=-\nabla_x E_y^{\mathrm I}(x)=Ax+\gamma b_y,
  \label{eq:ode}
\end{equation}
so the energy descends,
\begin{equation}
  \frac{d}{dt}E_y^{\mathrm I}(x(t))=-\|\nabla_x E_y^{\mathrm I}(x(t))\|^2\leq 0.
  \label{eq:lyapunov}
\end{equation}
Each spin is clipped at $\pm1$, as a node voltage saturates at the supply rails of an analog Ising machine. This only removes terms from Eq.~\eqref{eq:lyapunov}, so $E_y^{\mathrm I}$ remains non-increasing. The decoder reads $x(T)$ at a fixed horizon $T$ and produces the final image.

\subsection{Kuramoto Oscillator Dynamics}
\label{sec:kuramoto}

The Kuramoto oscillator model uses $n$ coupled phases $\theta\in\mathbb{T}^n$~\citep{kuramoto1975self}. We follow the oscillator-based Ising machine formulation~\citep{wang2019oim}. Starting from independent random uniform phases, the oscillator dynamics are
\begin{equation}
  \theta_i(0)\sim\mathcal{U}([0,2\pi)), \qquad
  \dot{\theta}_i = \sum_j K_{ij}\sin(\theta_j-\theta_i)- b_{y,i}\sin\theta_i,
  \label{eq:kuramoto}
\end{equation}
where $K=K^\top\in\mathbb{R}^{n\times n}$ is the learned, class-shared coupling and $b_y\in\mathbb{R}^n$ is the local conditioning vector, which absorbs the strength $\gamma$. The final term couples oscillator $i$ to a reference phase fixed at zero. Since $K$ is symmetric, Eq.~\eqref{eq:kuramoto} is the descent of a Lyapunov energy,
\begin{equation}
  E_y^{\mathrm K}(\theta) = -\frac{1}{2}\sum_{i\neq j}K_{ij}\cos(\theta_j-\theta_i)
  -\sum_i b_{y,i}\cos\theta_i,
  \label{eq:xy}
\end{equation}
because $\dot{\theta}=-\nabla_\theta E_y^{\mathrm K}$. Consequently,
\begin{equation}
  \frac{d}{dt}E_y^{\mathrm K}(\theta(t))=-\|\nabla_\theta E_y^{\mathrm K}(\theta(t))\|^2\leq 0.
  \label{eq:descent}
\end{equation}
Oscillator-based Ising machines perform this descent for optimization problems, and we apply it to generation. We stop the dynamics at a fixed horizon $T$. The final state $[\cos\theta(T),\sin\theta(T)]$ is fed to the decoder.

\subsection{Local Conditioning Mechanism}
\label{sec:tilt}

\begin{figure}[t!]
    \centering
    \includegraphics[width=0.95\linewidth]{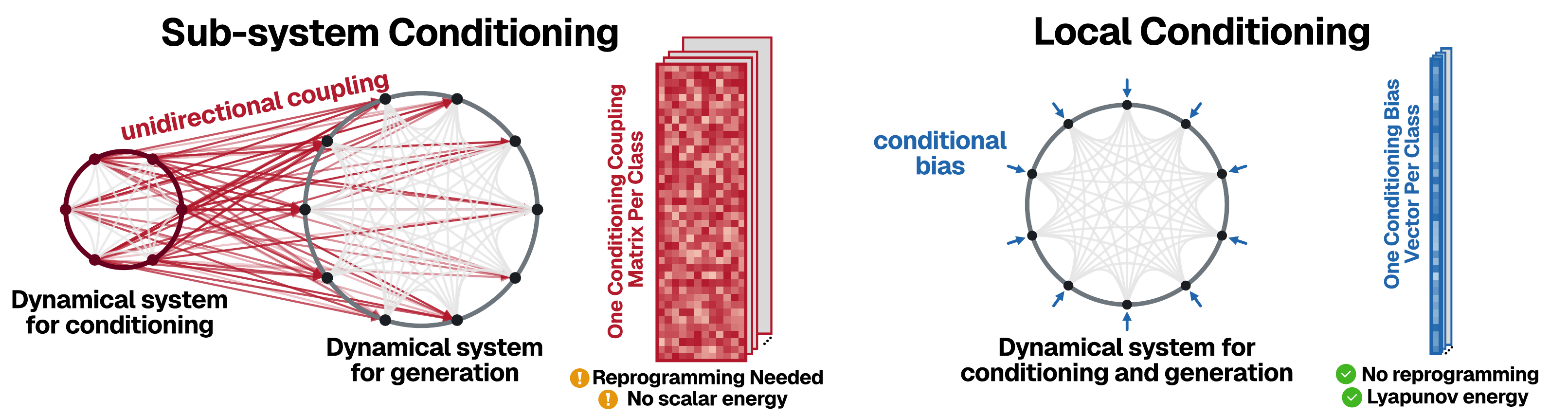}
    \caption{Comparison of conditioning models. Sub-system conditioning adds an auxiliary system and class-specific unidirectional couplings. Each class needs its own coupling, so reprogramming is required, and the dynamics generally do not descend a scalar energy. Our approach instead applies one bias per node for each class, keeping a fixed pairwise interaction and a Lyapunov energy.}
    \label{fig:conditioning-architectures}
\end{figure}

\begin{wrapfigure}{r}{0.47\textwidth}
    \centering
    \includegraphics[width=0.47\textwidth]{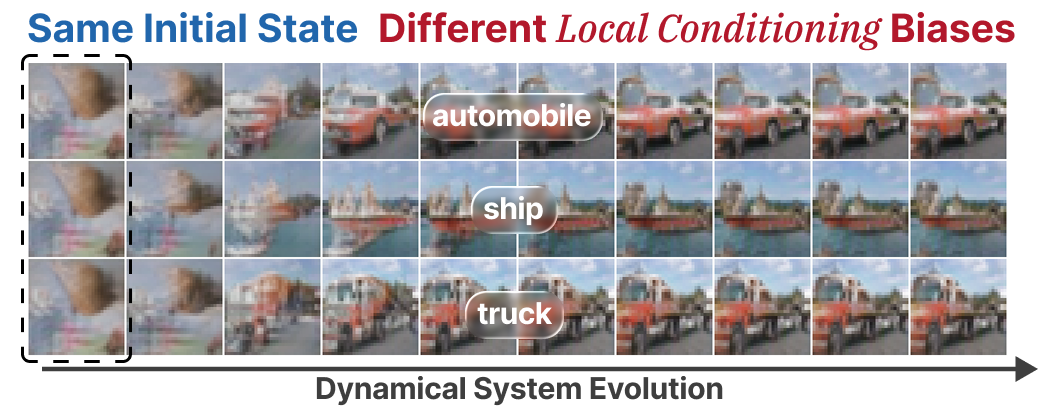}
    \caption{One random initial state evolves under different local conditioning biases, producing images of different classes.}
    \label{fig:conditioning}
\end{wrapfigure}

One naive way to build conditional generative models from hardware primitives is to train one model per class. On a physical substrate, this choice is costly: each model has its own coupling matrix, so one model per class means either one chip per class or rewriting the full interaction in the hardware (e.g., memristive or photonic arrays) at every class switch. Prior work~\citep{unconventional_ai_un0_2026} reduces the changes to a subset of the interactions (Figure~\ref{fig:conditioning-architectures}, left). It evolves a separate group of oscillators alongside the $n$ oscillators for generation and connects the two systems with a class-dependent, unidirectional coupling. This design has several drawbacks: it adds dynamical states, requires coupling reprogramming, and generally lacks the energy-descent property favored by many hardware substrates. Reprogramming changes what the substrate was built to hold fixed and adds recalibration cost. This leaves a gap between models and hardware.

We close this gap by moving the class conditioning out of the interaction and into the local term of each node (Figure~\ref{fig:conditioning-architectures}, right). The pairwise interaction is programmed once and never changed, and no auxiliary system is added. Selecting class $y$ injects a term into each node $i$: the bias $\gamma b_{y,i}$ in the Ising model and the reference coupling $b_{y,i}$ in the Kuramoto model. The mechanism has a direct interpretation: the interaction defines an energy landscape shared by all classes, and the conditioning vector tilts it. Every class descends the same landscape under a different tilt, so the Lyapunov property, the programmed hardware, and its calibration are all reused. The label enters the energies of Eq.~\eqref{eq:energy} and Eq.~\eqref{eq:xy} only through the local term. For the same state, $x$ or $\theta$, and two labels $y$ and $y'$, the shared interaction $A$ or $K$ cancels in the energy difference:
\begin{equation}
  \begin{aligned}
    E_y^{\mathrm I}(x)-E_{y'}^{\mathrm I}(x)
    &=-\gamma(b_y-b_{y'})^\top x,\\
    E_y^{\mathrm K}(\theta)-E_{y'}^{\mathrm K}(\theta)
    &=-\sum_i(b_{y,i}-b_{y',i})\cos\theta_i.
  \end{aligned}
  \label{eq:split-kuramoto}
\end{equation}
Only the biases remain. Figure~\ref{fig:conditioning} shows the effect of the conditioning mechanism. Starting from one random initial state, we run the dynamics with the biases of three classes: automobile, ship, and truck. As the dynamical-system state evolves, its decoded image settles to the selected class.

\subsection{Training}
\label{sec:loss}
We focus on efficient inference on the substrate. Model training is performed only once on conventional hardware. We train the dynamics and the decoder jointly using an objective that aligns generated and real feature distributions of each class in frozen DINOv2 feature spaces~\citep{oquab2024dinov2}, which encourages both image fidelity and within-class diversity. Following the drifting objective~\citep{deng2026generativemodelingdrifting}, we match generated features toward targets constructed from Sinkhorn displacements with a self-matching correction~\citep{he2026sinkhorn}, motivated by the debiased Sinkhorn divergence~\citep{genevay2018learning,feydy2019interpolating}. More details are provided in Appendix~\ref{app:training}.

\subsection{Non-von Neumann Hardware Realization}
\label{sec:hw_mapping}

Our dynamics build on primitives found in many non-von Neumann substrates: relaxation, programmable pairwise couplings, and per-node biases. We take Ising machines as an example. For the Kuramoto realization, oscillator-based Ising machines implement the pairwise term of Eq.~\eqref{eq:kuramoto}. Coupling each oscillator to a shared reference adds the local conditioning mechanism, as CMOS oscillator chips already do~\citep{lo2023ising}. For the Ising realization, resistively coupled electronic Ising machines~\citep{liu2025integrated} store $x$ as capacitor voltages, and programmable couplers inject currents proportional to $Ax$ that charge each capacitor. Per-node bias inputs supply $\gamma b_y$ for local conditioning. Section~\ref{sec:case-study} details this design. Furthermore, photonic~\citep{inagaki2016coherent,kalinin2025analog}, compute-in-memory~\citep{cai2020power,jiang2023efficient}, thermodynamic~\citep{melanson2025thermodynamic}, and superconducting~\citep{razmkhah2024josephson} substrates already offer most of these primitives, making them natural candidates as well.

\section{Evaluation}
\label{sec:eval}

Because the proposed model is not tied to a single physical substrate, we separate model-level capability from implementation-level performance. At the model level (Section~\ref{sec:eval-generation}), we compare against substrate-native models. At the implementation level (Section~\ref{sec:conventional}), we pair our model with a concrete hardware design and compare its energy and latency against conventional models on GPUs.

\subsection{Model-Level Capability}
\label{sec:eval-generation}
\vspace{-5pt}
\begin{table}[t]
    \centering
    \caption{Results on CIFAR-10, grouped by system size $n$. $\dagger$ denotes reported results. Our rows show the best of three seeds. \textbf{Reprogram}: requires reprogramming at each class switch. \textbf{IS}: Inception Score.}
    \vspace{6pt}
    \label{tab:results}
    \renewcommand{\arraystretch}{0.9}
    \begin{tabular*}{\textwidth}{@{\extracolsep{\fill}}lcrrrr@{}}
        \toprule
        \textbf{Model} & \textbf{Reprogram} & \textbf{Total Params} &
        \textbf{Decoder Params} & \textbf{IS $\uparrow$} &
        \textbf{clean-FID $\downarrow$} \\
        \midrule
        Un-0.n1024$^\dagger$ & Yes & 1.29M  & 0.16M & 9.84 & 11.16 \\
        Ising-1024            & No  & 0.60M  & 0.07M & 9.32 & 11.57 \\
        Kuramoto-1024         & No  & 0.69M  & 0.16M & 8.99 & \textbf{10.86} \\
        \midrule
        Un-0.n2048$^\dagger$ & Yes & 4.94M  & 0.58M & 10.23 & 9.38 \\
        Ising-2048            & No  & 2.28M  & 0.16M & 10.01 & \textbf{7.62} \\
        Kuramoto-2048         & No  & 2.70M  & 0.58M & 9.94 & 8.26 \\
        \midrule
        Un-0.n4096$^\dagger$ & Yes & 19.43M & 2.33M & 10.40 & 8.86 \\
        Ising-4096            & No  & 9.01M  & 0.58M & 10.12 & 6.40 \\
        Kuramoto-4096         & No  & 10.75M & 2.33M & 10.34 & \textbf{5.37} \\
        \bottomrule
    \end{tabular*}
\end{table}

Prior non-von Neumann substrates for generative tasks have mainly been evaluated on simple benchmarks such as MNIST~\citep{lecun1998mnist}. On CIFAR-10~\citep{krizhevsky2009cifar}, the closest baseline is Un-0~\citep{unconventional_ai_un0_2026}, which uses coupled oscillators with varied frequencies and changes the cross-system interactions for class conditioning. We compare the proposed Ising and Kuramoto models with Un-0 at system sizes $n\in\{1024,2048,4096\}$.

\paragraph{Results.}
Table~\ref{tab:results} reports the comparison. Regarding image generation quality, our models match the baseline at $n=1024$, and the margin grows with system size. At $n=4096$, both lower clean-FID~\citep{parmar2021cleanfid} by more than $2$ while keeping a competitive Inception Score (IS)~\citep{IS}. In terms of modeling, our Kuramoto dynamics do not require varied natural frequencies, which eases hardware implementation, and both proposed models descend an explicit Lyapunov energy. For conditioning, the baseline reprograms cross-system couplings at every class switch, whereas our local conditioning updates only one bias per node. These benefits come without sacrificing image quality, which even improves at larger $n$. The proposed approach thus narrows the gap between generative models and hardware realization and requires only minor modifications to existing substrates.

\subsection{Implementation-Level Performance}
\label{sec:conventional}

To evaluate implementation-level efficiency, we pair the proposed model with a concrete hardware design (Section~\ref{sec:case-study}) and compare its performance with models running on GPUs. Our data are projected from post-layout transistor-level simulation results of a CMOS-compatible dynamical system. Appendix~\ref{sec:circuit-sim} provides the circuit details.

\begin{figure}[!t]
    \centering
    \includegraphics[width=1\linewidth]{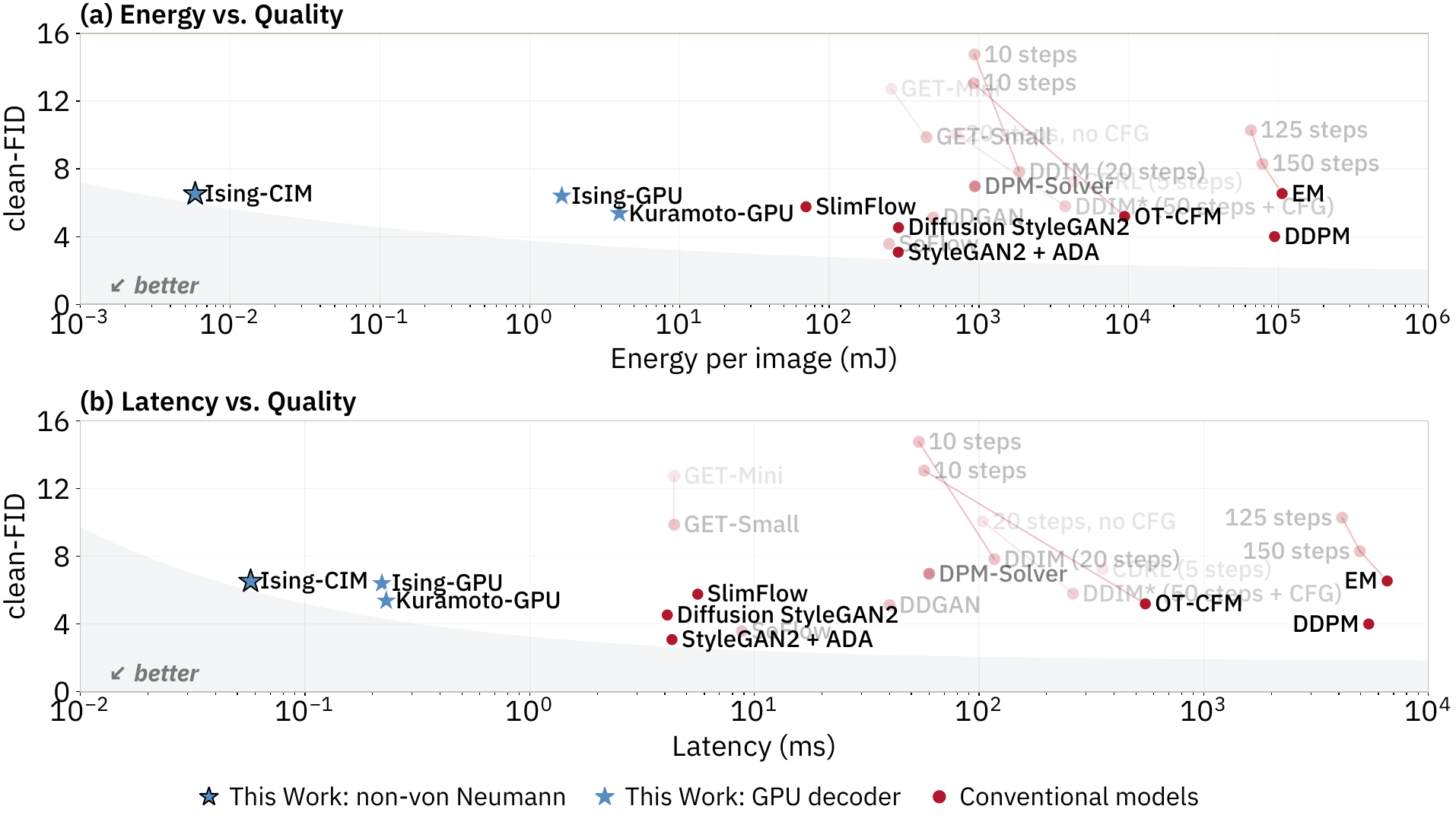}
    \caption{(a) Energy per generated image vs. clean-FID. (b) Latency vs. clean-FID. Some data points are faded for better visualization.}
    \label{fig:impl-perf}
\end{figure}

\paragraph{Baselines.}
We compare against the conventional generators shown in Figure~\ref{fig:impl-perf}: DDPM~\citep{ho2020ddpm}, DDIM~\citep{song2021ddim}, DPM-Solver~\citep{lu2022dpmsolver}, OT-CFM~\citep{tong2024otcfm}, CDRL~\citep{zhu2024cdrl}, Energy Matching~\citep{balcerak2025energymatching}, DDGAN~\citep{xiao2022ddgan}, SlimFlow~\citep{zhu2024slimflow}, SoFlow~\citep{luo2026soflow}, GET~\citep{geng2023get}, Diffusion-StyleGAN2~\citep{wang2023diffusiongan}, and StyleGAN2-ADA~\citep{karras2020ada}. We additionally train a class-conditional DDPM U-Net whose parameter count approximately matches that of Ising-4096 and sample it with DDIM (DDIM*). All GPU measurements use an NVIDIA L40S GPU.

\paragraph{Our models.}  Based on projections from circuit simulations (Section~\ref{sec:case-study} and Appendix~\ref{app:core-cost}), the Ising-4096 core draws $1.37$--$8.05$~W and is read out after $10.1$--$80.8$~ns, costing ${\sim}110.6$~nJ per image. Relaxation costs orders of magnitude less than GPU inference. The reason is physical: the substrate itself carries out the computation, with no clocked arithmetic and no data movement between memory and processor~\citep{zhang2024reviewising}. The Kuramoto model behaves similarly (Appendix~\ref{app:core-cost}). Local conditioning maintains this advantage across classes. The compact decoder is far smaller than every baseline model and can run on GPUs or dedicated accelerators.

\paragraph{Results.}
Figure~\ref{fig:impl-perf}(a) compares the energy cost per generated image against clean-FID. When the decoder is mapped onto compute-in-memory (CIM) hardware, the Ising-CIM (4096 spins) is located at the extreme low-energy end of the Pareto frontier, requiring only 5.9~\textmu J per image. Under decoder quantization constraints, it achieves a clean-FID of $6.52$. With the decoder executed on a GPU, the resulting Ising-GPU and Kuramoto-GPU (4096 nodes) realizations consume energy at the millijoule scale, still orders of magnitude less than most GPU baselines. For the latency comparison in Figure~\ref{fig:impl-perf}(b), our implementations operate at sub-millisecond latency, more than an order of magnitude faster than the GPU baselines. Overall, our models deliver strong generation quality with substantially lower energy and latency than GPU-based generators.

\subsection{Ablation Study}
\label{sec:ablation}

To measure what the learned dynamical system adds beyond the decoder, we compare the full Ising and Kuramoto models at $n=4096$ with two variants. \textbf{Reservoir} keeps the dynamical-system evolution but freezes the coupling matrix at its random initialization, and only the local conditioning vectors and decoder are trained. \textbf{Decoder Only} removes the dynamical system and feeds the decoder a random state with a learned class-specific offset, preserving conditional generation.

\begin{wraptable}{r}{0.5\textwidth}
    \centering
    \vspace{-15pt}
    \caption{Contribution of the dynamical system to image generation quality. Mean clean-FID ($\downarrow$) over three seeds.}
    \label{tab:ablation}
    \setlength{\tabcolsep}{3pt}
    \vspace{6pt}
    \begin{tabular}{lrr}
        \toprule
        \textbf{Variant} & \textbf{Kuramoto-4096} & \textbf{Ising-4096} \\
        \midrule
        Full         & 5.41  & 6.53  \\
        Reservoir    & 10.02 & 13.96 \\
        Decoder Only & 10.14 & 18.19 \\
        \bottomrule
    \end{tabular}
\end{wraptable}
For each model, all variants use the same decoder architecture, per-seed initialization of common components, and the same full training schedule. We report the mean clean-FID over three random seeds. Table~\ref{tab:ablation} shows that the learned dynamical system improves generation quality in both families. \textbf{Reservoir} and \textbf{Decoder Only} increase clean-FID over the full models by $4.6$ to $11.7$, with larger gaps for Ising, even though these variants still train their conditioning vectors and decoders.

\section{Case Study: A CMOS-Compatible Dynamical System}
\label{sec:case-study}
\subsection{Hardware Design}

\begin{figure}[b]
    \centering
    \includegraphics[width=\linewidth]{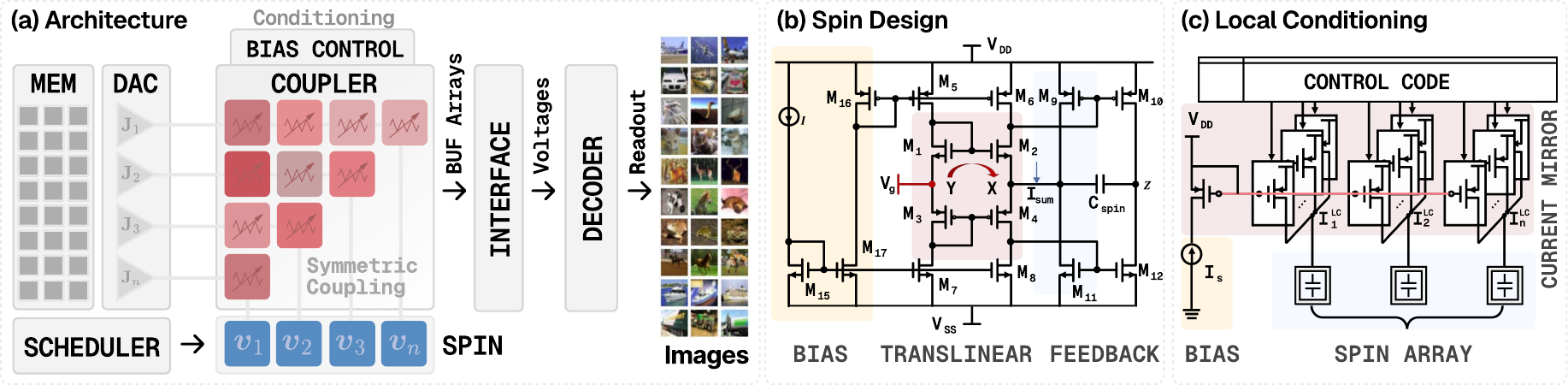}
    \caption{CMOS-compatible dynamical-system implementation. (a) Overall architecture. (b) Circuit design for spins. (c) Circuit design for local conditioning.}
    \label{fig:cmos-implementation}
\end{figure}

\paragraph{System architecture.} The proposed principle applies to many emerging hardware substrates. As a concrete example, we map the Ising model onto the core of a CMOS-compatible Ising machine~\citep{liu2025integrated}. As shown in Figure~\ref{fig:cmos-implementation}(a), programmable resistive couplers connect the spins. The analog voltage on each spin capacitor drives this network through a unity-gain buffer. The system state evolves continuously. To support the local conditioning mechanism, a digitally controlled bias current is applied to each spin. The hardware therefore encodes the energy defined in Eq.~\eqref{eq:energy}. A scheduler coordinates the operation. After initialization, the system starts to evolve along an energy landscape defined by the coupling coefficients and reshaped by the bias currents. Finally, the continuous-spin voltages are read out, buffered, and sent to the decoder to render the image.

\paragraph{Conditioning.} A key advantage is that the coupling coefficients stay fixed across image-class requests. The coupler array is therefore programmed and calibrated only once.  For each request, the corresponding bias currents are set in parallel. For a fully connected 4096-spin Ising substrate with symmetric 5-bit couplings, switching to a different image class requires changing only the conditioning currents. No reprogramming or recalibration follows.

\paragraph{Circuit design.} The continuous-spin circuit in Figure~\ref{fig:cmos-implementation}(b) uses a current conveyor that holds the coupling node $X$ near the reference voltage $V_{\mathrm{g}}$ while conveying the bidirectional coupling current $I_{\mathrm{sum}}$ to the spin capacitor $C_{\mathrm{spin}}$. The capacitor integrates the net coupling current, and its voltage represents the continuous-spin state. Figure~\ref{fig:cmos-implementation}(c) shows the local conditioning circuit, where a reference current $I_S$ biases a current-mirror array and locally stored binary control codes independently set the bias current applied to each spin.

\paragraph{Decoder.}

The decoder is a seven-layer convolutional network that requires 26M, 34M, and 86M multiply-accumulates (MACs) per generated image for $n=1024$, $2048$, and $4096$, respectively. Its convolution operations can be mapped to existing mixed-signal and CIM accelerators as matrix-vector multiplications with locally stored weights~\citep{shafiee2016isaac,sebastian2020memory}. For a hardware-based energy estimate, we use results from fabricated CIM chips: a 55-nm 3D-RRAM CIM reports 29.94~TOPS/W for 4-bit inputs and 5-bit weights~\citep{huo2022rram}, while a 55-nm Twin-8T SRAM-CIM reports 37.5~TOPS/W for 2-bit inputs and 5-bit weights~\citep{si2020twin8t}. Counting two operations per MAC, these measured efficiencies give estimated decoder energies of 1.39--1.74~\textmu J, 1.81--2.27~\textmu J, and 4.59--5.74~\textmu J per image for $n=1024$, $2048$, and $4096$, respectively. For latency, a fabricated 16-nm multibit CIM accelerator reports a measured peak MAC throughput of 3.0~TOPS at 8-bit precision and 11.8~TOPS at 4-bit precision, and demonstrates an end-to-end throughput of 7{,}815~images/s for a fully mapped 11-layer 4-bit CNN~\citep{jia2022scalable}. Using the measured 3.0-TOPS point as a conservative throughput reference gives compute-only latency estimates of approximately 17.3~\textmu s, 22.7~\textmu s, and 57.3~\textmu s for the three decoder sizes.

\paragraph{Validation.} To validate the proposed architecture at the circuit level, we perform post-layout transistor-level simulations of the dynamical-system core in a 65-nm CMOS technology using Cadence Virtuoso AMS with Spectre as the analog simulator. A 64-spin model for MNIST generation serves as the test case (Appendix~\ref{sec:circuit-sim}). The continuous-spin states are read out after 10.1~ns, a time set by the model dynamics and the circuit RC time constant. The simulated energy consumption of the dynamical-system core is 0.216~nJ per image.

\subsection{Mitigating Hardware Non-Idealities}
\label{sec:pvt}
\begin{table}[t]
    \centering
    \caption{Robustness of the proposed models to analog hardware non-idealities at $n=4096$. Every stored weight of the dynamical system is quantized to $k$ bits and read back with element-wise multiplicative Gaussian noise.}
    \vspace{5pt}
    \label{tab:hardware}
    \setlength{\tabcolsep}{3.8pt}
    \renewcommand{\arraystretch}{0.9}
    \begin{tabular}{lccccccccc}
        \toprule
        \textbf{Weight precision} & \multicolumn{3}{c}{\textbf{Full precision}} & \multicolumn{3}{c}{\textbf{5-bit quantized}} & \multicolumn{3}{c}{\textbf{4-bit quantized}} \\
        \cmidrule(lr){2-4} \cmidrule(lr){5-7} \cmidrule(lr){8-10}
        \textbf{Noise level} & None & 2\% & 5\% & None & 2\% & 5\% & None & 2\% & 5\% \\
        \midrule
        \multicolumn{10}{l}{\textit{Kuramoto-4096}} \\
        clean-FID $\downarrow$ & 5.37 & 5.39 & 5.27 & 5.24 & 5.34 & 5.13 & 5.37 & 5.30 & 5.34 \\
        Change vs.\ ideal & -- & $+0.02$ & $-0.10$ & $-0.13$ & $-0.03$ & $-0.24$ & $0.00$ & $-0.07$ & $-0.03$ \\
        \midrule
        \multicolumn{10}{l}{\textit{Ising-4096}} \\
        clean-FID $\downarrow$ & 6.40 & 6.72 & 6.66 & 6.40 & 6.33 & 6.64 & 6.56 & 6.87 & 6.87 \\
        Change vs.\ ideal & -- & $+0.32$ & $+0.26$ & $0.00$ & $-0.07$ & $+0.24$ & $+0.16$ & $+0.47$ & $+0.47$ \\
        \bottomrule
    \end{tabular}
\end{table}

Emerging hardware substrates, such as analog computing platforms, often suffer from non-idealities~\citep{xiao2022accuracy}, including limited programming precision, circuit variations, and noise. We therefore train and evaluate the dynamics under these effects. Every stored weight, namely the coupling $A$ or $K$ and the local conditioning vectors $b_y$, is quantized to $k$ bits by a quantizer $Q_k$ scaled to the tensor's largest magnitude, and read back as $W_{\mathrm{hw}}=Q_k(W)\odot(\mathbf{1}+\sigma\varepsilon)$ with $\varepsilon\sim\mathcal{N}(0,I)$ and noise level $\sigma\in\{0,2\%,5\%\}$. During training, gradients pass through the rounding via a straight-through estimator~\citep{bengio2013estimating}. Table~\ref{tab:hardware} shows that both models tolerate these non-idealities. In every tested setting, clean-FID increases by less than $0.5$. With quantization- and noise-aware training, the models therefore operate under realistic hardware constraints at little cost in generation quality.

\section{Related Work}
\label{sec:related}
\paragraph{Hopfield networks and Boltzmann machines.}
Hopfield networks connect recurrent dynamics to an energy landscape~\citep{hopfield1982}, and the graded-response extension carries energy descent over to continuous states~\citep{hopfield1984}. Their purpose is associative memory: stored patterns are attractors, and a corrupted cue evolves to the matching fixed point. Modern Hopfield networks raise memory capacity and relate their update rule to transformer attention~\citep{ramsauer2020hopfield,krotov2023new,hu2023sparse,NEURIPS2025_634ddc25}, and the Energy Transformer minimizes a global energy with recurrent attention~\citep{hoover2023energy}. Boltzmann machines use a similar pairwise energy to define an equilibrium distribution, $p(x)\propto\exp[-E(x)]$, and learn by matching equilibrium statistics~\citep{ackley1985}, often approximated by contrastive divergence~\citep{hinton2002contrastive}. Our models share this pairwise energy but use it differently. Our energy landscape does not define a likelihood, and generation does not require reaching equilibrium. Instead, $E_y^{\mathrm I}$ and $E_y^{\mathrm K}$ specify a descent that turns a random state into a representation for the decoder, and training optimizes the resulting samples.

\paragraph{Generative models.}
Diffusion models generate samples by reversing a prescribed forward noising process using a learned time-dependent denoising or score function~\citep{sohl2015deep,ho2020ddpm,yang2023diffusion}. Flow matching~\citep{liu2022flow,lipman2023flow,davis2024fisher} instead learns time-conditioned velocities that transport a base distribution along a predefined probability path. Both explicitly supervise dynamics in data or latent space during training. Equilibrium Matching~\citep{wang2025eqm} learns a time-invariant gradient over noise--data interpolations so that sampling descends a fixed energy landscape. Our energy has a different role: it is defined over the physical system's internal state space, and a decoder maps the final states to images. It neither models the image density nor supplies a supervised transport. Conditioning is a local term for each node that tilts a shared energy landscape. The generator has no forward noising process or learned time-dependent transport.

\paragraph{Ising machines and dynamical systems.}
Ising machines were first built for combinatorial optimization: they map a problem to an Ising energy and search for low-energy states.
Implementations include photonic coherent Ising machines~\citep{inagaki2016coherent,yamamoto2017coherent,hamerly2019experimental}, oscillator-based Ising machines~\citep{chou2019analog,wang2019oim,vaidya2022creating,moy2022ring,albertsson2023highly,cilasun2025coupled} and their theoretical models~\citep{erementchouk2022computational,cdc9-y234}, and CMOS~\citep{afoakwa2021brim,su2023reconfigurable,wang2024energy,liu2025integrated,liu2025ising} and superconducting~\citep{razmkhah2024josephson,ucpinar2024scalable} circuits.
They have been applied to Max-Cut~\citep{mcmahon2016fully,haribara2016coherent,wang2019oim,bashar2020experimental,ucpinar2024scalable,shaglel2026comprehensive}, SAT~\citep{bashar2023designing,bybee2023efficient,sharma2023augmenting,su2023reconfigurable,sikhakollu2024high,de2025incorporate,sun2025general}, MIMO detection~\citep{singh2022ising,sreedhara2023mu}, and other optimization problems~\citep{tanahashi2019application,matsumoto2022distance,parizy2022cardinality,bao2023ising,kirihara2023exploring,mao2023chemical,tsuyumine2024optimization}. Related Ising-derived dynamical systems have been trained for graph learning, function learning, and differential equation solving~\citep{song2024ds,wu2024extending,liu2025instatrain,liu2025expressive,liu2025ds}.

\section{Conclusion}

This work presents a design principle for efficient image generation on non-von Neumann hardware. We develop continuous Ising and Kuramoto models that turn physical energy minimization into generative computation. Both models descend explicit energy functions and map naturally to existing hardware. For conditional generation, we close the gap between models and hardware with local conditioning: a local term for each node selects the class while the programmed interactions stay fixed. This is attractive for physical systems such as memristive and photonic arrays, where writing and calibrating pairwise couplings costs more than updating biases. Our evaluation shows that high-quality image generation can emerge from a different and restricted computational primitive, with projected efficiency far beyond that of GPUs.

Several questions remain open. Richer energy functions may make the model more expressive but need more complex hardware. Our current implementation is fully connected: exploring sparse or structured connectivity could substantially reduce coupler, programming, and storage requirements. Finally, validating the models on fabricated hardware under measured noise, precision, programming, and calibration constraints is the next step toward the hardware-efficiency goal of this work.

\newpage

\bibliography{ref}
\bibliographystyle{unsrtnat}

\newpage
\appendix
\section{Training Objective}
\label{app:training}
We jointly train the dynamics and the decoder by matching generated and real image distributions within each class. This distribution-level supervision encourages both image fidelity and within-class diversity. We perform the matching in frozen DINOv2 feature spaces~\citep{oquab2024dinov2}, so that the loss reflects semantic content and local structure rather than pixelwise similarity alone.

\paragraph{Feature regression.}
Following the drifting objective~\citep{deng2026generativemodelingdrifting}, for each class $y$ and feature view $F_v$, let $g_i=F_v(G(y,\xi_i))$ denote the feature of generated sample $i$. We compute a displacement $V_i$ that specifies how $g_i$ should change to better match the real feature distribution. Its normalized form $\widehat V_i$ defines a desired step in feature space. We use the shifted feature as a temporary regression target:
\begin{equation}
\tau_i=\operatorname{sg}\!\left(g_i+\widehat V_i\right), \qquad \ell_i=\|g_i-\tau_i\|_2^2.
\end{equation}
The target represents a desired new feature position, not a paired real image. The stop-gradient operator $\operatorname{sg}$ holds $\tau_i$ fixed during backpropagation, yielding $\nabla_{g_i}\ell_i=-2\widehat V_i$. We average the losses over samples, classes, and feature views.

\paragraph{Sinkhorn displacement.}
To construct $V_i$, we use Sinkhorn-based distribution matching~\citep{he2026sinkhorn} to align generated features $\{g_i\}$ with same-class real references $\{r_j\}$, favoring nearby features while balancing assignments across all real references. Each generated feature is guided toward a weighted average of its matched references. This averaging, however, can pull distinct features together even when the generated and real distributions already agree. To correct this effect, we also match the current generated batch to an independent batch from the same generator and class, which estimates the attraction that soft matching introduces within the generated distribution itself. Let $M_{\mathrm{real}}(g_i)$ and $M_{\mathrm{self}}(g_i)$ denote the weighted reference means obtained from Sinkhorn matching to real and generated references, respectively. We construct
\begin{equation}
V_i=\bigl(M_{\mathrm{real}}(g_i)-g_i\bigr)-\bigl(M_{\mathrm{self}}(g_i)-g_i\bigr)
=M_{\mathrm{real}}(g_i)-M_{\mathrm{self}}(g_i).
\end{equation}
The first term points toward real data, and the second subtracts the self-attraction estimated from generated data. This correction is motivated by the debiased Sinkhorn divergence~\citep{genevay2018learning,feydy2019interpolating}.

\section{Conditioning for Asymmetric Coupling}
Local conditioning also applies to asymmetric coupling: one bias per node still selects the class, although the dynamics are no longer guaranteed to descend the energies. Asymmetric coupling may not be available on some hardware substrates. We evaluate this setting in Table~\ref{tab:results-asym}. The dynamical system then requires roughly twice as many parameters as in the symmetric case.

\begin{table}[h]
    \centering
    \caption{Asymmetric Ising and Kuramoto models on CIFAR-10, grouped by system size $n$.}
    \label{tab:results-asym}
    \begin{tabular}{lcrrr}
        \toprule
        \textbf{Model} & \textbf{\begin{tabular}[b]{@{}c@{}}Reprogramming\\Needed\end{tabular}} & \textbf{\begin{tabular}[b]{@{}c@{}}Total\\Params\end{tabular}} & \textbf{\begin{tabular}[b]{@{}c@{}}Decoder\\Params\end{tabular}} & \textbf{clean-FID $\downarrow$} \\
        \midrule
        Ising-1024-asym & No & 1.13M & 0.07M & 9.71 \\
        Kuramoto-1024-asym & No & 1.22M & 0.16M & 14.15 \\
        \midrule
        Ising-2048-asym & No & 4.37M & 0.16M & 7.15 \\
        Kuramoto-2048-asym & No &  4.80M & 0.58M & 8.23 \\
        \midrule
        Ising-4096-asym & No & 17.40M & 0.58M & 5.96 \\
        Kuramoto-4096-asym & No & 19.14M & 2.33M & 5.39 \\
        \bottomrule
    \end{tabular}
\end{table}

\section{Dynamical-System Core Cost}
\label{app:core-cost}

Table~\ref{tab:core-cost} summarizes the power and energy scaling of the continuous-spin dynamical-system core. The 64-spin post-layout simulation of Section~\ref{sec:case-study} serves as the reference. For larger systems, the spin capacitance is conservatively scaled as $250~\mathrm{fF}\sqrt{n/64}$ to account for the increased aggregate coupling current. Two representative coupling scenarios are considered. Case~A approximately keeps the coupling density and resistance distribution of the reference network, yielding stronger aggregate coupling and a similar evolution time as the system scales. Case~B instead holds a roughly constant number of nonzero couplings per spin, reducing coupling power at the cost of a longer evolution time. These cases represent two practical scaling scenarios rather than strict bounds. The same projection method is applied to the Kuramoto oscillator systems. From 64 to 128 oscillators, the simulated energy per image increases from $0.396$ to $0.832~\mathrm{nJ}$, with the evolution time increasing from $170$ to $180~\mathrm{ns}$. For larger systems, we conservatively assume that the coupling power scales with $n^2$ and the bias power scales with $n$. The projected energy per image is $10.32$, $27.71$, and $84.22~\mathrm{nJ}$ for 1024, 2048, and 4096 oscillators, respectively.

\begin{table}[t]
\centering
\caption{Power and energy per image of the dynamical-system core with $R_0=40~\mathrm{k}\Omega$. The Reference row is obtained from post-layout transistor-level simulation, while Cases A and B are projections. $P_{\mathrm{spin}}$, $P_{\mathrm{coup}}$, and $P_{\mathrm{bias}}$ denote the power of the continuous-spin, coupler, and bias circuits, respectively. $P_{\mathrm{tot}}$ is their sum, and $t_{\mathrm{read}}$ is the evolution time.}
\label{tab:core-cost}

\footnotesize
\setlength{\tabcolsep}{4.5pt}

\begin{tabular}{c c c c c c c c c}
\toprule
\textbf{\boldmath$n$} & \textbf{\boldmath$C_{\mathrm{spin}}$} & \textbf{Case} & \textbf{\boldmath$P_{\mathrm{spin}}$} & \textbf{\boldmath$P_{\mathrm{coup}}$} & \textbf{\boldmath$P_{\mathrm{bias}}$} & \textbf{\boldmath$P_{\mathrm{tot}}$} & \textbf{\boldmath$t_{\mathrm{read}}$} & \textbf{\boldmath$E/\mathrm{image}$} \\
\midrule
64 & 250 fF & Reference & 19.20 mW & 1.657 mW & 0.538 mW & 21.40 mW & 10.1 ns & 0.216 nJ \\
\midrule
128 & 353.6 fF & A & 38.40 mW & 6.628 mW & 1.076 mW & 46.10 mW & 10.1 ns & 0.466 nJ \\
128 & 353.6 fF & B & 38.40 mW & 3.314 mW & 1.076 mW & 42.79 mW & 14.28 ns & 0.611 nJ \\
\midrule
1024 & 1.000 pF & A & 307.2 mW & 424.2 mW & 8.608 mW & 740.0 mW & 10.1 ns & 7.47 nJ \\
1024 & 1.000 pF & B & 307.2 mW & 26.51 mW & 8.608 mW & 342.3 mW & 40.4 ns & 13.83 nJ \\
\midrule
2048 & 1.414 pF & A & 614.4 mW & 1.697 W & 17.22 mW & 2.329 W & 10.1 ns & 23.52 nJ \\
2048 & 1.414 pF & B & 614.4 mW & 53.02 mW & 17.22 mW & 684.6 mW & 57.13 ns & 39.11 nJ \\
\midrule
4096 & 2.000 pF & A & 1.229 W & 6.787 W & 34.43 mW & 8.050 W & 10.1 ns & 81.31 nJ \\
4096 & 2.000 pF & B & 1.229 W & 106.0 mW & 34.43 mW & 1.369 W & 80.8 ns & 110.6 nJ \\
\bottomrule
\end{tabular}
\end{table}

\section{Circuit Simulation}
\label{sec:circuit-sim}
\subsection{Circuit-to-Dynamics Correspondence}

In our circuit implementation, each continuous spin is represented by the voltage $v_j$ stored on $C_{\mathrm{spin}}$ relative to the fixed reference voltage $V_{\mathrm{g}}$. A differential amplifier generates the two polarities $V_{\mathrm{g}}+v_j$ and $V_{\mathrm{g}}-v_j$, such that the sign of each coupling coefficient is implemented by selecting the corresponding polarity at the input of the current-conveyor-based coupling path. The magnitude of the dimensionless coefficient $A_{ij}$ is mapped to the physical conductance through a normalization resistance $R_0$. With the local bias current $I_{b,j}$ represented equivalently by $b_j=R_0I_{b,j}$, the relevant circuit quantities and the resulting spin dynamics are
\begin{equation}
\begin{aligned}
v_j &= V_j-V_{\mathrm{g}}, \qquad G_{ij}=\frac{|A_{ij}|}{R_0}, \qquad b_j=R_0I_{b,j}, \\
\frac{dv_j}{dt} &= \frac{dV_j}{dt}
=\frac{1}{R_0C_{\mathrm{spin}}}\left(\sum_i A_{ij}v_i+b_j\right).
\end{aligned}
\label{eq:circuit_dynamics}
\end{equation}
Here, the selection of the positive or negative spin polarity gives the signed coupling contribution $A_{ij}v_i/R_0$, while the programmed conductance determines its magnitude.

For the symmetric interaction matrix $A_{ij}=A_{ji}$, the circuit dynamics descend the energy
\begin{equation}
E_{\mathrm c}(v)=-\frac{1}{2}\sum_j\sum_i A_{ij}v_i v_j-\sum_j b_j v_j
=-\frac{1}{2}v^\top A v-b^\top v,
\label{eq:circuit_potential}
\end{equation}
for which
\begin{equation}
-\frac{\partial E_{\mathrm c}}{\partial v_j}=\sum_i A_{ij}v_i+b_j, \qquad
\frac{dv_j}{dt}=-\frac{1}{R_0C_{\mathrm{spin}}}\frac{\partial E_{\mathrm c}}{\partial v_j}.
\label{eq:gradient_dynamics}
\end{equation}
The continuous-spin network therefore follows the dynamics in the unsaturated operating region. Along its state trajectory, the energy satisfies
\begin{equation}
\frac{dE_{\mathrm c}}{dt}
=\sum_j\frac{\partial E_{\mathrm c}}{\partial v_j}\frac{dv_j}{dt}
=-R_0C_{\mathrm{spin}}\sum_j\left(\frac{dv_j}{dt}\right)^2\leq 0,
\label{eq:potential_descent}
\end{equation}
showing descent of $E_{\mathrm c}(v)$ under the circuit dynamics. The physical voltage range of the spin nodes bounds the feasible state space, while the programmed resistive coupling network determines the energy landscape through $A$. With $v=V_sx$ for a voltage scale $V_s$ and $b_j=\gamma V_s b_{y,j}$, we have $E_{\mathrm c}(v)=V_s^2E_y^{\mathrm I}(x)$, and Eq.~\eqref{eq:circuit_dynamics} becomes Eq.~\eqref{eq:ode} with time unit $R_0C_{\mathrm{spin}}$.

\subsection{Transient Mixed-Signal Operation}

We perform a post-layout transient simulation of the hardware implementation in Section~\ref{sec:case-study}. The simulation consists of initialization, autonomous evolution, and readout. During initialization, the prescribed continuous-spin states are loaded onto the spin capacitors while the coupling coefficients and local conditioning currents are configured. The initialization paths are then disconnected simultaneously, after which the network evolves continuously through the interaction of the couplers, current-conveyor interfaces, conditioning currents, and spin capacitors. The analog solver integrates these circuit dynamics, whereas the event-driven digital domain is used only to sequence initialization, configuration, and readout. At readout, all continuous-spin voltages are simultaneously transferred to the hold capacitors through unity-gain input buffers. After the sampling switches are opened, the held voltages are isolated from the continuously evolving state nodes and buffered for subsequent processing by the image decoder.

\begin{figure}[t!]
    \centering
    \includegraphics[width=\linewidth]{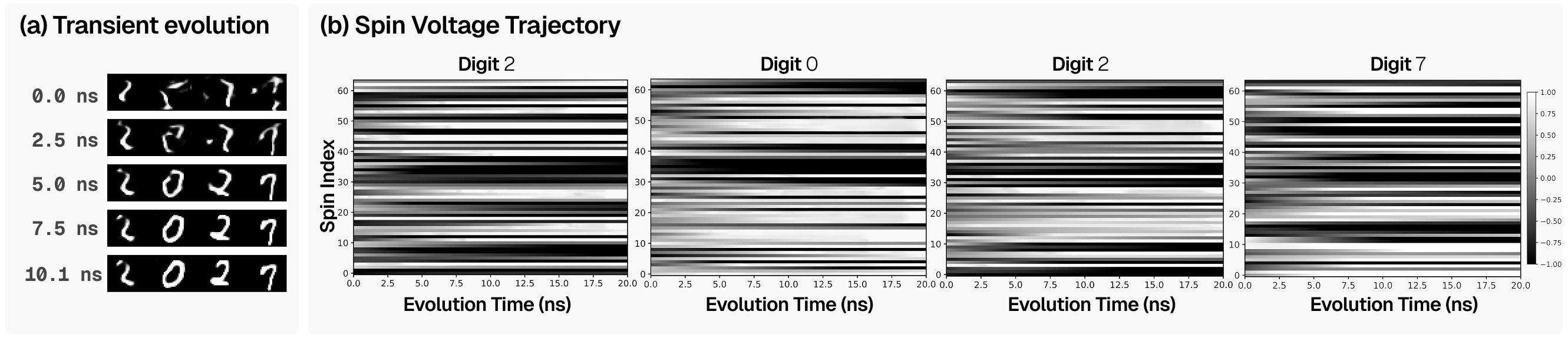}
    \caption{(a) Transient evolution of the decoded ``2027'' image. (b) Circuit-level spin-voltage trajectories of the four samples over the first 20~ns. The final spin states are sampled at 10.1~ns.}
    \label{fig:transient_evolution}
\end{figure}

All 64 continuous-spin trajectories are recorded throughout the circuit evolution. Figure~\ref{fig:transient_evolution}(b) shows the circuit-level spin trajectories of the four samples that generate the digits ``2027'' over the first 20~ns. To visualize the corresponding evolution in the image domain, the circuit-extracted latent states are passed through the trained decoder at five representative time points, as shown in Figure~\ref{fig:transient_evolution}(a). The decoded images progressively approach the target digit patterns, with the continuous-spin states sampled for final readout at 10.1~ns.

\subsection{Large-Scale Circuit Validation}
\begin{wrapfigure}{r}{0.45\textwidth}
    \centering
    \includegraphics[width=0.45\textwidth]{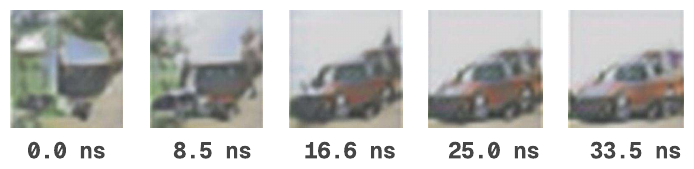}
    \caption{Evolution of one generated  \textit{automobile} image from a $1024$-spin circuit simulation.}
    \vspace{-1pt}
    \label{fig:Cifar_ising}
\end{wrapfigure}
Although the preceding results use post-layout transistor-level simulation, extending the same Cadence setup to larger systems becomes computationally demanding. For the large-scale simulation, selected circuit blocks are therefore replaced by ideal elements parameterized from their corresponding circuit implementations, while the network dynamics and coupling topology are unchanged. This model enables simulation of a 1024-spin system in Cadence for CIFAR-10 image generation. Figure~\ref{fig:Cifar_ising} shows the evolution of one generated image. The observed evolution times fall within the range estimated in Table~\ref{tab:core-cost}, providing a consistency check between the large-scale simulation and the circuit-based performance projection.

\end{document}